\documentclass{emulateapj}
\usepackage{apjfonts}

\usepackage{amsmath}

\usepackage{graphics}

\newcommand{\bd}{\begin{displaymath}}
\newcommand{\ed}{\end{displaymath}}

\slugcomment{accepted by ApJ}

\shorttitle{On the disappearance of BLR in LLAGNs}

\begin{document}

\title{On the disappearance of broad-line region in low-luminosity active galactic nuclei:
the role of the outflows from advection dominated accretion flows}

\author{Xinwu Cao\altaffilmark{1}}

\begin{abstract}
The broad-line region (BLR) disappears in many low-luminosity active
galactic nuclei (AGNs), the reason of which is still controversial.
The BLRs in AGNs are believed to be associated with the outflows
from the accretion disks. Most of the low-luminosity AGNs (LLAGNs)
contain advection dominated accretion flows (ADAFs), which are very
hot and have a positive Bernoulli parameter. ADAFs are therefore
associated with strong outflows. We estimate the cooling of the
outflows from the ADAFs, and find that the gases in such hot
outflows always cannot be cooled efficiently by bremsstrahlung
radiation. The ADAF may co-exist with the standard disk, i.e., the
inner ADAF connects to the outer thin accretion disk at radius
$R_{\rm d,tr}$, in the sources accreting at slightly lower than the
critical rate $\dot{m}_{\rm crit}$ ($\dot{m}=\dot{M}/\dot{M}_{\rm
Edd}$). For the ADAFs with $L_{\rm bol}/L_{\rm Edd}\ga 0.001$, a
secondary small inner cold disk is suggested to co-exist with the
ADAF due to the condensation process. We estimate the Compton
cooling of the outflow, of which the soft seed photons either come
from the outer cold disk or the secondary inner cold disk. It is
found that the gas in the outflow far from the ADAF may be
efficiently cooled to form BLR clouds due to the soft seed photons
emitted from the cold disks, provided the transition radius of the
ADAF to the outer cold disk is small [$r_{\rm d,tr}=R_{\rm
d,tr}/(2GM/c^2)\la 20$] or/and the secondary small cold disk has a
luminosity $L_{\rm sd}\ga 0.003L_{\rm Edd}$.  The BLR clouds can
still be formed in the outflows from the outer cold thin disks, if
the transition radius $r_{\rm tr}$ is not very large. For the
sources with $L_{\rm bol}/L_{\rm Edd}\la 0.001$, the inner small
cold disk is evaporated completely in the ADAF and outer thin
accretion disk may be suppressed by the ADAF, which leads to the
disappearance of the BLR. The physical implications of this scenario
on the double-peaked broad-line emitters are also discussed.


\end{abstract}


\altaffiltext{1}{Key Laboratory for Research in Galaxies and
Cosmology, Shanghai Astronomical Observatory, Chinese Academy of
Sciences, 80 Nandan Road, Shanghai, 200030, China; cxw@shao.ac.cn}

\section{Introduction}

Active galactic nuclei (AGNs) are classified as type 1 and 2 AGNs by
their line emission. Type 1 AGNs show broad emission lines and
narrow forbidden lines, while only narrow lines are observed in type
2 AGNs. According to the unification scheme of AGNs, all AGNs are
intrinsically same, but are viewed at different orientations
\citep*[e.g.,][]{1993ARA&A..31..473A}. The broad-line regions (BLRs)
in type 2 AGNs are obscured by the dusty tori, as they are supposed
to be viewed at large angles with respect to the axes of the tori.
However, there is evidence that the BLR disappears in many
low-luminosity active galactic nuclei (LLAGNs)
\citep*[e.g.,][]{2001ApJ...554L..19T,2003ApJ...583..632T,2002ApJ...579..205G},
and most of the type 1 AGNs have relatively high Eddington ratios
\citep*[e.g.,][]{2009ApJ...700...49T}. These low-luminosity sources
are named as ``true" type 2 AGNs, which do not have hidden BLRs
\citep*[see][for a review and references
therein]{2008ARA&A..46..475H}. Many workers have explored why the
BLR disappears in LLAGNs
\citep*[e.g.,][]{2003ApJ...589L..13N,2003ApJ...590...86L,2006ApJ...648L.101E,2009ApJ...701L..91E}.
\citet{2003ApJ...590...86L} suggested that an upper limit on the
observed width of broad emission lines leads to a lower limit on the
radius of the BLR based on the empirical correlation between BLR
size and optical continuum luminosity \citep{2000ApJ...533..631K}.
In this scenario, the BLR radius shrinks below a critical value for
LLAGNs, which leads to the disappearance of BLR in these sources.
Although the origin of BLR is still unclear, an attractive
suggestion is that the BLR structure is associated with the outflow
from the accretion disk \citep{1992ApJ...385..460E}.
\citet{2000ApJ...530L..65N} assumed that the winds from the
accretion disk are triggered by the thermal instability of radiation
pressure dominated region of the disk \citep{1976MNRAS.175..613S}.
The transition radius between the radiation pressure dominated and
gas pressure dominated regions in the disk increases with the
dimensionless mass accretion rate $\dot{m}$
\citep{1973A&A....24..337S}. In this scenario, the transition radius
becomes smaller than the marginal stable orbit of the black hole for
low accretion rates (low luminosities), and the winds are switched
off and no BLR can be formed in LLAGNs \citep{2003ApJ...589L..13N}.
A correlation between the width of BLR and the luminosity is
expected in this model, which is consistent with the observations of
AGN samples \citep{2004ApJ...608..136W,2007ChJAA...7...63X}. An
alternative disk-wind scenario was suggested for the BLR and dust
torus, in which both the BLR and torus disappear when the bolometric
luminosity is low \citep{2006ApJ...648L.101E,2009ApJ...701L..91E}.
The outflow from the accretion disk being switched off is a key
ingredient in these scenarios when accretion rates are low, though
the detailed physics of the outflow dynamics has not been included
in these works.

Low mass accretion rate $\dot{m}$ may lead to the accretion flows to
be advection-dominated
\citep{1994ApJ...428L..13N,1995ApJ...452..710N}. Advection dominated
accretion flows (ADAFs) are suggested to be present in LLAGNs
\citep*[see][for a review and references
therein]{2002luml.conf..405N}, which can successfully explain most
observational features of LLAGNs
\citep*[e.g.,][]{1996ApJ...462..142L,1999ApJ...516..177G,1999ApJ...525L..89Q,2009RAA.....9..401X}.
{It was suggested that the ADAF co-exists with the standard disk,
i.e., the inner ADAF connects to the outer thin accretion disk, in
some sources accreting at rates slightly lower than the critical
rate $\dot{m}_{\rm crit}$
\citep*[e.g.,][]{1997ApJ...489..865E,1999ApJ...525L..89Q}. For even
lower accretion rates, a secondary small cold accretion disk is
suggested to co-exist with the ADAF in the inner region due to the
condensation process \citep{2000A&A...360.1170R}. This model was
extensively explored by many different authors
\citep*[e.g.,][]{2007A&A...463....1M,2007ApJ...671..695L,2007MNRAS.376..435M,2008ApJ...688..527T},
which can explain the soft X-ray thermal component observed in some
X-ray binaries \citep{2008ApJ...680..593T,2006ApJ...652L.113M}.}
\citet{2004A&A...428...39C} assumed that the existence of the BLR is
related with the cold accretion disk, and they compared different
theoretical model predictions with the observations of AGNs, which
favors the disappearance of BLR being related with the different
accretion mode in LLAGNs. Based on the evaporation disk model, a
lower limit on the accretion rate is also derived for the existence
of BLRs on the same assumption that the BLR is associated with a
cold accretion disk \citep{2009ApJ...707..233L}.

It was well known that the gas in ADAFs is very hot, and has a
positive Bernoulli parameter, which implies that ADAFs should be
associated with strong winds
\citep{1994ApJ...428L..13N,1995ApJ...444..231N,1999MNRAS.303L...1B,1999MNRAS.310.1002S,2000ApJ...537L..27I,2001MNRAS.322..461S}.
If this is the case, one may expect strong outflows from the ADAFs
in LLAGNs, which implies that the assumption of the disk winds being
suppressed at low accretion rates in the previous scenarios for the
disappearance of BLR in LLAGNs is not valid
\citep{2003ApJ...589L..13N,2009ApJ...701L..91E}. In this work, we
explore the relation of the hot outflows from ADAFs with the
disappearance of BLR in LLAGNs.

\section{Cooling of the outflows from ADAFs in LLAGNs}

In this work, we assume that the outflow has a conical geometry, and
the density of the outflow from an ADAF is
\begin{equation}
\rho(R)={\frac {\dot{M}_{\rm w}}{f_{\rm w}R^2v(R)}}, \label{dens_w}
\end{equation}
where $\dot{M}_{\rm w}$ is the mass loss rate in the outflow, $v(R)$
is the radial velocity of the outflow at radius $R$, and $f_{\rm w}$
is the solid angle of the conical outflow ($f_{\rm w}=4\pi$ for an
isotropic outflow). The mass loss rate $\dot{M}_{\rm w}$ in the
outflow is related to the mass accretion rate $\dot{M}$ of the disk
with
\begin{equation}
\dot{M}_{\rm w}=\eta_{\rm w}\dot{M}, \label{eta_w}
\end{equation}
where $\eta_{\rm w}$ is a free parameter and required to be less
than unity. For the outflow driven by the internal energy of the hot
gases in the ADAF, its velocity
\begin{equation}
v(R)\sim \left({\frac {GM}{R}}\right)^{1/2}, \label{v_w}
\end{equation}
where $M$ is the mass of the black hole. In principle, the velocity
can be higher than this value, which is the least velocity of the
outflow can escape to infinity.

Substituting Eqs. (\ref{eta_w}) and (\ref{v_w}) into Eq.
(\ref{dens_w}), we have
\begin{equation}
\rho(r)=7.51\times 10^{-4}\eta_{\rm w}f_{\rm
w}^{-1}\dot{m}m^{-1}r^{-3/2}~ {\rm g~cm^{-1}}, \label{dens_w2}
\end{equation}
where the dimensionless quantities are defined as
\begin{equation}
m={\frac {M}{M_\odot}},~~r={\frac {R}{2GM/c^2}},~~\dot{m}={\frac
{\dot{M}}{\dot{M}_{\rm Edd}}},\label{quan_dml}
\end{equation}
and
\begin{displaymath}
\dot{M}_{\rm Edd}=1.39\times 10^{18}m~{\rm g~s^{-1}}.
\end{displaymath}
 The temperature of the gases in the ADAFs is nearly virialized
 \citep{1995ApJ...444..231N}, and we assume the gases to be virialized in the
 outflow,
\begin{equation}
T_{\rm gas}(R)\sim T_{\rm vir}(R)={\frac {GMm_{\rm p}}{kR}}.
\label{t_gas}
\end{equation}
The internal energy per unit volume of the gases in the outflow is
\begin{equation}
U={\frac {3}{2}}p_{\rm gas}={\frac {3\rho kT_{\rm i}}{2\mu_{\rm
i}m_{\rm p}}}+{\frac {3\rho kT_{\rm e}}{2\mu_{\rm e}m_{\rm p}}},
\label{inter_eng}
\end{equation}
where the effective molecular weights of the ions and electrons are
$\mu_{\rm i}=1.23$ and $\mu_{\rm e}=1.14$ respectively. As the ion
temperature is significantly higher than the electron temperature in
the inner region of the ADAF and most of the internal energy is
stored in the ions, the electron temperature $T_{\rm e}\le T_{\rm
i}$ is required in the outflow. The electron temperature $T_{\rm e}$
is mainly determined by the radiative cooling, and the Coulomb
interaction between the electrons and ions. In this work, we assume
$T_{\rm e}=\xi_{\rm e}T_{\rm gas}$ ($\xi_{\rm e}\le 1$) in our
estimates on the cooling of the outflow. Thus, the bremsstrahlung
cooling timescale of the gases in the outflow can be estimated as
\begin{equation}
\tau_{\rm cool}^{\rm brem}\sim {\frac {U}{F_{\rm brem}^{-}}},
\label{tau_cool}
\end{equation}
where the bremsstrahlung cooling rate in unit volume of the gases is
\citep{1986rpa..book.....R}
\begin{equation}
F_{\rm brem}^{-}=2.36\times 10^{-27}n_{\rm e}^2T_{\rm e}^{1/2}~{\rm
erg~s^{-1} cm^{-3}}. \label{f_minus}
\end{equation}
Substituting Eqs. (\ref{t_gas}), (\ref{inter_eng}) and
(\ref{f_minus}) into Eq. (\ref{tau_cool}), the bremsstrahlung
cooling timescale of the gases in the outflow is available,
\begin{equation}
\tau_{\rm cool}^{\rm brem}(r)\sim {\frac {U}{F_{\rm
brem}^{-}}}=1.00\times 10^{-3}f_{\rm w}\eta_{\rm w}^{-1}\xi_{\rm
e}^{-1/2}m\dot{m}^{-1}r~{\rm s}. \label{tau_cool2}
\end{equation}
The cooling length scale of the outflow is therefore estimated by
\begin{equation}
l_{\rm cool}(r)=\tau_{\rm cool}v=2.12\times 10^{7}f_{\rm w}
\eta_{\rm w}^{-1}\xi_{\rm e}^{-1/2}m\dot{m}^{-1}r^{1/2}~{\rm cm}.
\label{l_cool}
\end{equation}
The mass accretion rate $\dot{m}$ being lower than a critical value
$\dot{m}_{\rm crit}$ is required for an ADAF. The critical rate
$\dot{m}_{\rm crit}\simeq 0.01$ is suggested either by the
observations or the theoretical models \citep*[see][for a review and
references therein]{2002luml.conf..405N}. The lower limit on $l_{\rm
cool}(r)$ is derived as
\begin{equation}
l_{\rm cool}^{\rm min}(r)=\tau_{\rm cool}v=2.12\times 10^{9}f_{\rm
w} \eta_{\rm w}^{-1}m r^{1/2}~{\rm cm}, \label{l_cool_min}
\end{equation}
if $\dot{m}=\dot{m}_{\rm crit}=0.01$ and $\xi_{\rm e}=1$ are
substituted into Eq. (\ref{l_cool}), i.e., the electrons and ions
have the same temperature in the outflow. The electron temperature
should be significantly lower than the ion temperature at the base
of the outflow, because it comes from a two-temperature ADAF
\citep{1995ApJ...452..710N}. As the cooling rate increases with
electron temperature $T_{\rm e}$, the estimate performed with
$T_{\rm e}=T_{\rm i}$ gives the minimal cooling timescale (see Eq.
\ref{l_cool}). Comparing the cooling length scale with radius $R$,
we have
\begin{equation}
{\frac {l_{\rm cool}^{\rm min}(R)}{R}}=7.18\times 10^3 f_{\rm w}
\eta_{\rm w}^{-1}r^{-1/2}.  \label{l_cool_min2}
\end{equation}
The radiative cooling of the gases in the outflow is inefficient if
$l_{\rm cool}^{\rm min}(R)>R$, which leads to
\begin{equation}
r<5.15\times 10^7f_{\rm w}^2\eta_{\rm w}^{-2}. \label{l_cool_min3}
\end{equation}

The reverberation-mapping method
\citep{1997ASSL..218...85N,1993PASP..105..247P} was applied to
measure the size of the BLR from the time delay between the line and
continuum variations. The correlations between the optical
luminosity and BLR size were derived by different authors
\citep*[e.g.,][]{2000ApJ...533..631K,2006ApJ...644..133B}.
Subtracting the contribution from the host galaxy starlight to the
AGN emission, \citet{2006ApJ...644..133B} found that
\begin{equation}
\log R_{\rm BLR}=-21.69+0.518\log L_{\rm bol}, \label{r_blr}
\end{equation}
where $L_{\rm bol}\simeq 9\lambda L_\lambda (5100\AA)$ is used
\citep{2000ApJ...533..631K}. This is consistent with $R_{\rm
BLR}\propto L_{\rm bol}^{0.5}$ expected from the photoionization
model if all BLRs have similar physical properties. The distances
from the black hole in the outflow, within which the outflows are
radiatively cooled inefficiently (see Eq. \ref{l_cool_min3}), are
compared with the BLR sizes of broad-line AGNs in Fig.
\ref{fig_r_blr}. It is found that the radiative cooling is always
unimportant except in the region far from the BLRs, which implies
that the hot outflow from an ADAF is unable to be cooled to form BLR
clouds.

\vskip 1cm

\figurenum{1}
\centerline{\includegraphics[angle=0,width=8.5cm]{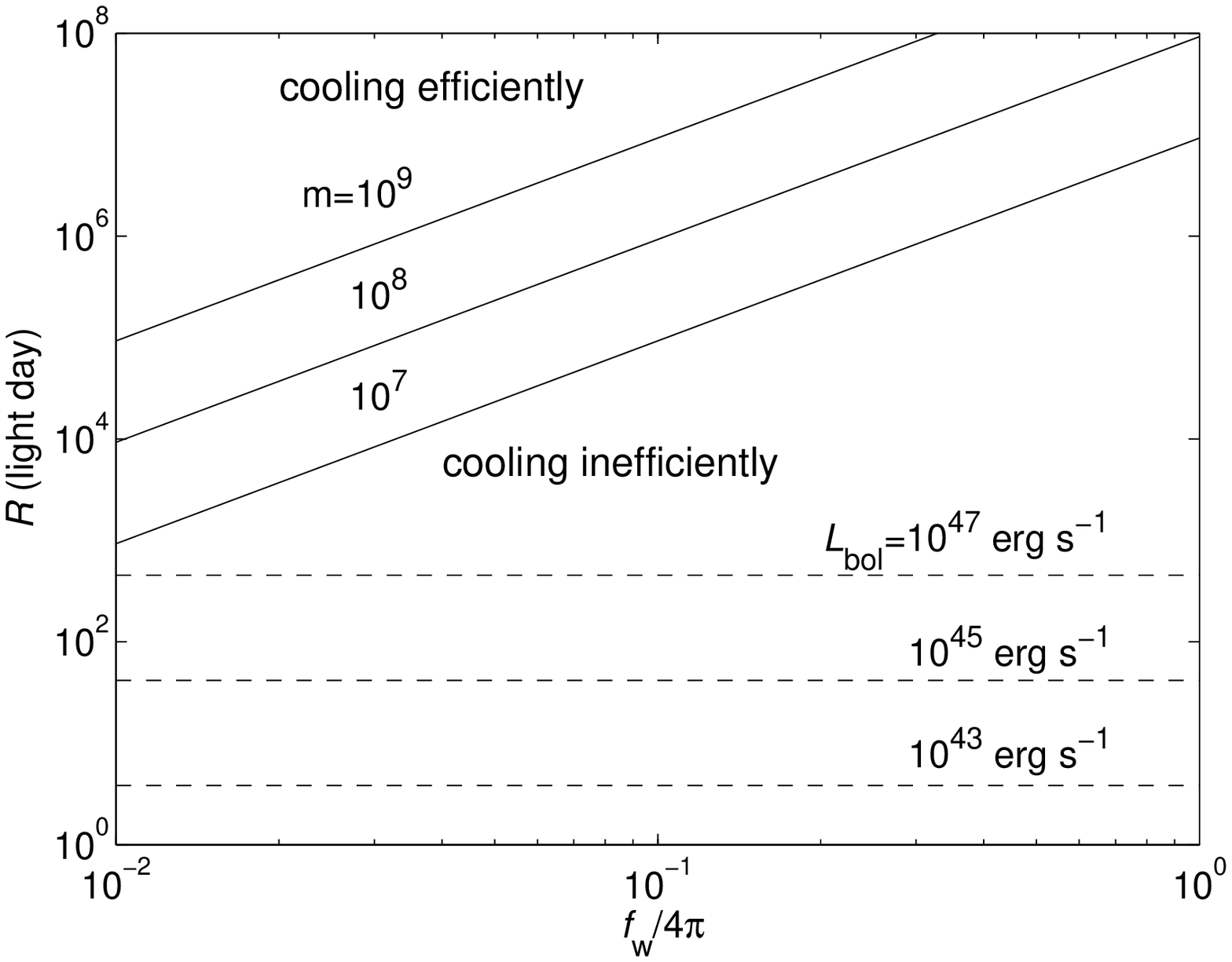}}
\figcaption{The distance from the black hole in the outflow (solid
lines), beyond which the outflow is bremsstrahlung cooled
efficiently, as functions of $f_{\rm w}$ with $\eta_{\rm w}=1$ (see
Eq. \ref{l_cool_min3}). For comparison, we also plot the BLR size
estimated from the bolometric luminosity with the empirical
correlation given by \citet{2006ApJ...644..133B} (dashed lines).
\label{fig_r_blr}   }\centerline{}

When the accretion rate is slightly lower than the critical value
$\dot{m}_{\rm crit}$, an ADAF is present near the black hole, and it
may connect to the outer standard disk at a transition radius
$R_{\rm d,tr}$. In this case, the soft photons emitted from the
outer cold disk will be Compton-upscattered by the hot electrons in
the outflow, and the plasma in the outflow is therefore cooled. The
flux due to viscous dissipation in the outer region of the disk is
\begin{equation}
F_{\rm vis}(R_{\rm d})\simeq {\frac {3GM\dot{M}}{8\pi R_{\rm d}^3}}.
\label{f_vis}
\end{equation}
The irradiation of the inner ADAF on the outer cold disk is almost
negligible compared with the viscous dissipation in the outer cold
disk, because the solid angle of the outer disk subtended to the
inner region of the ADAF is too small \citep{2006ApJ...652..112C}.
We neglect this effect in estimating the cooling caused by the
Compton scattering in the outflow. The cooling rate in unit volume
of the gases at radius $R$ in the outflow is
\begin{equation}
F_{\rm Comp}^{-}\simeq \int\limits_{R_{\rm d,tr}}^{R_{\rm
d,out}}{\frac {4kT_{\rm e}}{m_{\rm e}c^2}}{\frac {F_{\rm
vis}R}{\pi(R^2+R_{\rm d}^2)^{3/2}}}n_{\rm e}\sigma_{\rm T}2\pi
R_{\rm d}dR_{\rm d}, \label{f_comp_minus}
\end{equation}
where $n_{\rm e}$ is the number density of the electrons in the
outflow at $R$, and $\sigma_{\rm T}$ is the Thompson cross-section
of electron.
Using Eqs. (\ref{quan_dml}) and (\ref{t_gas}), we re-write Eq.
(\ref{f_comp_minus}) as
\begin{equation}
F_{\rm Comp}^{-}\simeq4.17\times10^6 \xi_{\rm e}n_{\rm
e}(r)m^{-1}\dot{m} \int\limits_{r_{\rm d,tr}}^{r_{\rm d,out}}{\frac
{dr_{\rm d}}{r_{\rm d}^2(r^2+r_{\rm d}^2)^{3/2}}}~{\rm erg~s^{-1}
cm^{-3}}, \label{f_comp_minus_b}
\end{equation}
where $r_{\rm d}=R_{\rm d}/(2GM/c^2)$. The Compton cooling timescale
for the outflow is available,
\begin{displaymath}
\tau_{\rm cool}^{\rm Comp}(r)\sim {\frac {U}{F_{\rm
Comp}^{-}}}=5.20\times 10^{-10}\xi_{\rm
e}^{-1}r^{-1}m\dot{m}^{-1}
\end{displaymath}
\begin{equation}
\times\left[\int\limits_{r_{\rm d,tr}}^{r_{\rm d,out}}{\frac
{dr_{\rm d}}{r_{\rm d}^2(r^2+r_{\rm d}^2)^{3/2}}}\right]^{-1}{\rm
s}, \label{tau_cool_comp}
\end{equation}
and the dynamical timescale of the outflow can be estimated by
\begin{equation}
\tau_{\rm dyn}\sim {\frac {R}{v}}={\frac
{R^{3/2}}{(GM)^{1/2}}}=1.39\times 10^{-5}mr^{3/2}~{\rm
s}.\label{tau_dyn}
\end{equation}
The importance of the Compton cooling of the gases in the outflow
can be evaluated by
\begin{equation}
{\frac {\tau_{\rm cool}^{\rm Comp}}{\tau_{\rm dyn}}}=3.74\times
10^{-5}\xi_{\rm e}^{-1}\dot{m}^{-1}r^{-5/2}\left[\int\limits_{r_{\rm
d,tr}}^{r_{\rm d,out}}{\frac {dr_{\rm d}}{r_{\rm d}^2(r^2+r_{\rm
d}^2)^{3/2}}}\right]^{-1}. \label{tau_ratio}
\end{equation}
{In the inner region of the ADAF, the electron temperature can be
more than one order of magnitude lower than the ion temperature
\citep*[][]{1995ApJ...452..710N}. Thus, the parameter $\xi_{\rm
e}\la 0.1$ in the base of the outflow from the ADAF, while $\xi_{\rm
e}\rightarrow 1$ in the outflow far from the black hole. } The
results derived with different disk parameters are plotted in Fig.
\ref{fig_comp}. We find that the timescale ratio, $\tau_{\rm
cool}^{\rm Comp}/\tau_{\rm dyn}$, decreases with increasing radius
$r$ in the outflow when $r$ is small (see Fig. \ref{fig_comp}),
because the solid angle of the outer cold disk region subtended to
the outflow increases with $r$ at small radii. At large radii, the
solid angle decreases with increasing $r$, and therefore the
timescale ratio, $\tau_{\rm cool}^{\rm Comp}/\tau_{\rm dyn}$,
increases with $r$. The Compton cooling becomes less important for a
disk accreting at a lower rate, because less soft seed photons are
emitted from the outer disk.

{For ADAFs in the sources with $L_{\rm bol}/L_{\rm Edd}\ga 0.001$, a
secondary small cold accretion disk extending to the marginal stable
orbit of the black hole can co-exist with an ADAF due the
condensation process. The outflow can be cooled due to the Compton
scattering of the soft seed photons emitted from such an inner cold
disk. The radiative power of the inner cold disk consists of the
viscously dissipated power in the disk and the power of the
irradiation from the ADAF. In order to avoid exploring the
complicated processes of the interaction between the ADAF and the
cold disk, we assume the flux from the unit surface area of the
inner cold disk to have the same radial dependence as the standard
cold disk \citep{1973A&A....24..337S},
\begin{equation}
F_{\rm vis}(R_{\rm d})= {\frac {\mathcal{C}_{\rm sd}mL_{\rm
sd}}{R_{\rm d}^3}} \left[1-\left({\frac{R_{\rm d,in}}{R_{\rm
d}}}\right)^{1/2}\right], \label{f_vis_smd}
\end{equation}
where $L_{\rm sd}$ is the luminosity of the small cold disk, and
$R_{\rm d,in}$ is the radius of the inner edge of the disk. This
small disk can extend to the marginal stable orbit of the black
hole, and we adopt $R_{\rm d,in}=R_{\rm d,ms}=6GM/c^2$ for a
non-rotating black hole in all our calculations. The luminosity of
the small disk is
\begin{equation}
L_{\rm sd}=2\int\limits_{R_{\rm d,min}}^{R_{\rm d,max}}F_{\rm
vis}(R_{\rm d})2\pi R_{\rm d}dR_{\rm d}, \label{l_smd}
\end{equation}
which leads to
\begin{displaymath}
\mathcal{C}_{\rm sd}=2.35\times 10^{4}\left\{ {\frac {1}{r_{\rm
d,min}}}\left[1-{\frac {2}{3}}\left({\frac {3}{r_{\rm
d,min}}}\right)^{1/2}\right]\right.
\end{displaymath}
\begin{equation}
\left.
 -{\frac {1}{r_{\rm
d,max}}}\left[1-{\frac {2}{3}}\left({\frac {3}{r_{\rm
d,max}}}\right)^{1/2}\right] \right\}^{-1}.\label{c_smd}
\end{equation}
Similar to the above estimates for the Compton cooling caused by the
emission from the outer cold disk, the ratio of the Compton cooling
timescale due to the presence of the inner small cold disk to the
dynamical timescale of the outflow is estimated as
\begin{equation}
{\frac {\tau_{\rm cool}^{\rm Comp}}{\tau_{\rm
dyn}}}=6.59\xi_{e}^{-1}\mathcal{C}_{\rm sd}^{-1}\lambda_{\rm
sd}^{-1}r^{-5/2}\left[\int\limits_{3}^{r_{\rm d,max}}{\frac
{[1-(3/r_{\rm d})^{1/2}]}{r_{\rm d}^2(r^2+r_{\rm
d}^2)^{3/2}}}dr_{\rm d}\right]^{-1}, \label{tau_ratio_smd}
\end{equation}
where the Eddington ratio of the small disk $\lambda_{\rm sd}=L_{\rm
sd}/L_{\rm Edd}$. The inner cold small disk is usually truncated at
several tens of Schwarzschild radii \citep{2007ApJ...671..695L}, and
$r_{\rm d,max}=20$ is therefore adopted in the estimates. The final
results are insensitive to the exact values of $r_{\rm d,max}$
adopted, because most of the emission is from the region of the disk
very close to the black hole. We plot the results in Fig.
\ref{fig_comp_smd}, which show that the Compton cooling of the
outflow near the ADAF due to the presence of the inner small
accretion disk is always unimportant, while the outflow can be
cooled efficiently at large distances from the black hole. }

\vskip 1cm

\figurenum{2}
\centerline{\includegraphics[angle=0,width=8.5cm]{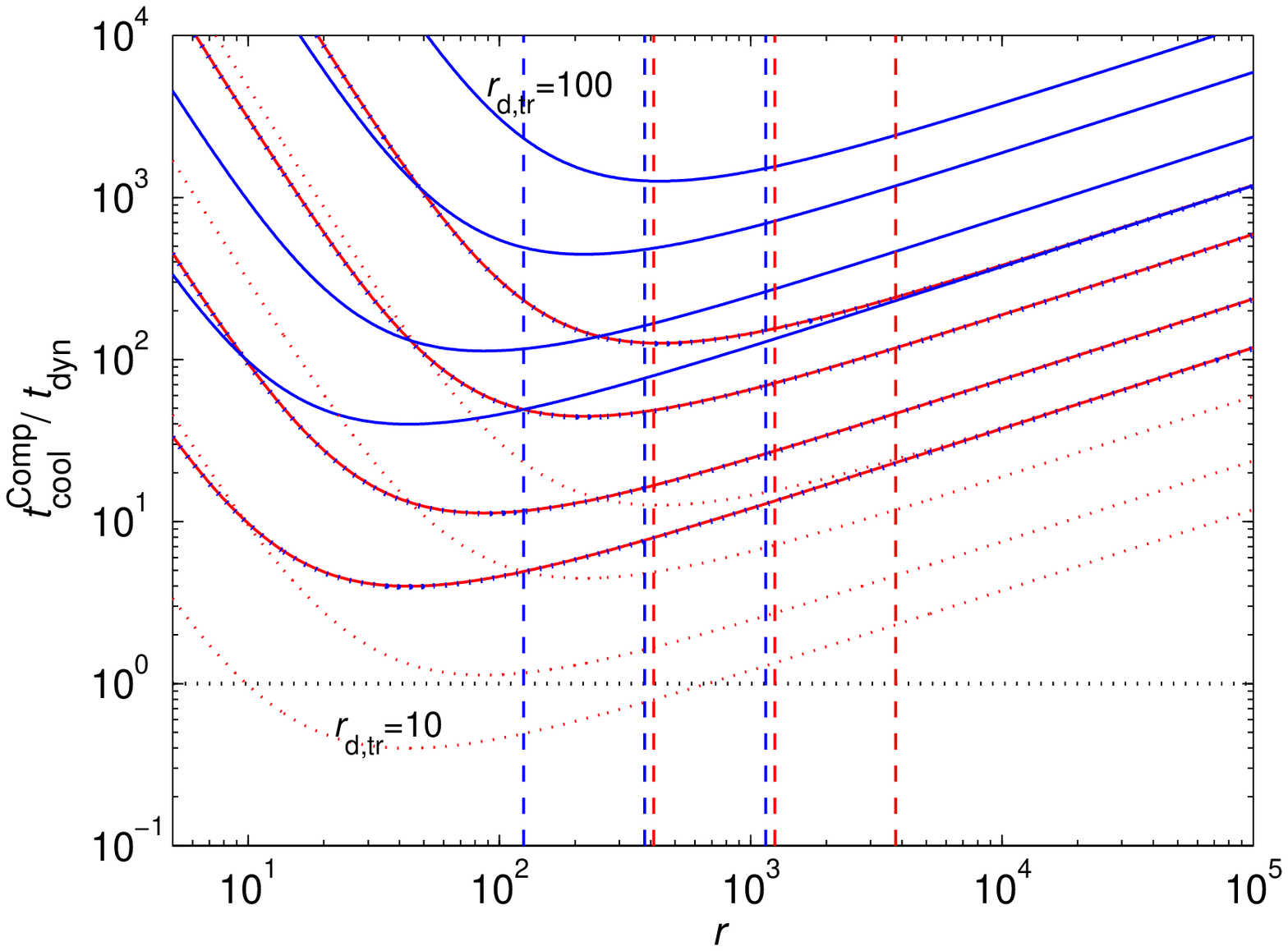}}
\figcaption{The ratios of the Compton cooling timescale to the
dynamical timescale as functions of radius $r$ for different model
parameters (solid and dotted color lines). The solid lines represent
the ratio calculated with $\xi_{\rm e}=0.1$ for different transition
radius, $r_{\rm tr}=10$, 20, 50, and 100, respectively (from bottom
to up), while the dotted color lines are the results calculated with
$\xi_{\rm e}=1$. The red lines are calculated with $\dot{m}=0.01$,
while $\dot{m}=0.001$ is adopted for the blue lines. For comparison,
we also plot the BLR size (dashed lines) estimated from the
bolometric luminosity with the empirical correlation given by
\citet{2006ApJ...644..133B} for different black hole mass, $m=10^7$,
$10^8$, and $10^9$, respectively (from right to left). The red
dashed lines correspond to the BLR sizes of AGNs with $L_{\rm
bol}/L_{\rm Edd}=0.01$, while the blue dashed lines are for $L_{\rm
bol}/L_{\rm Edd}=0.001$. \label{fig_comp}  }\centerline{}

 \vskip 1cm

\figurenum{3}
\centerline{\includegraphics[angle=0,width=8.5cm]{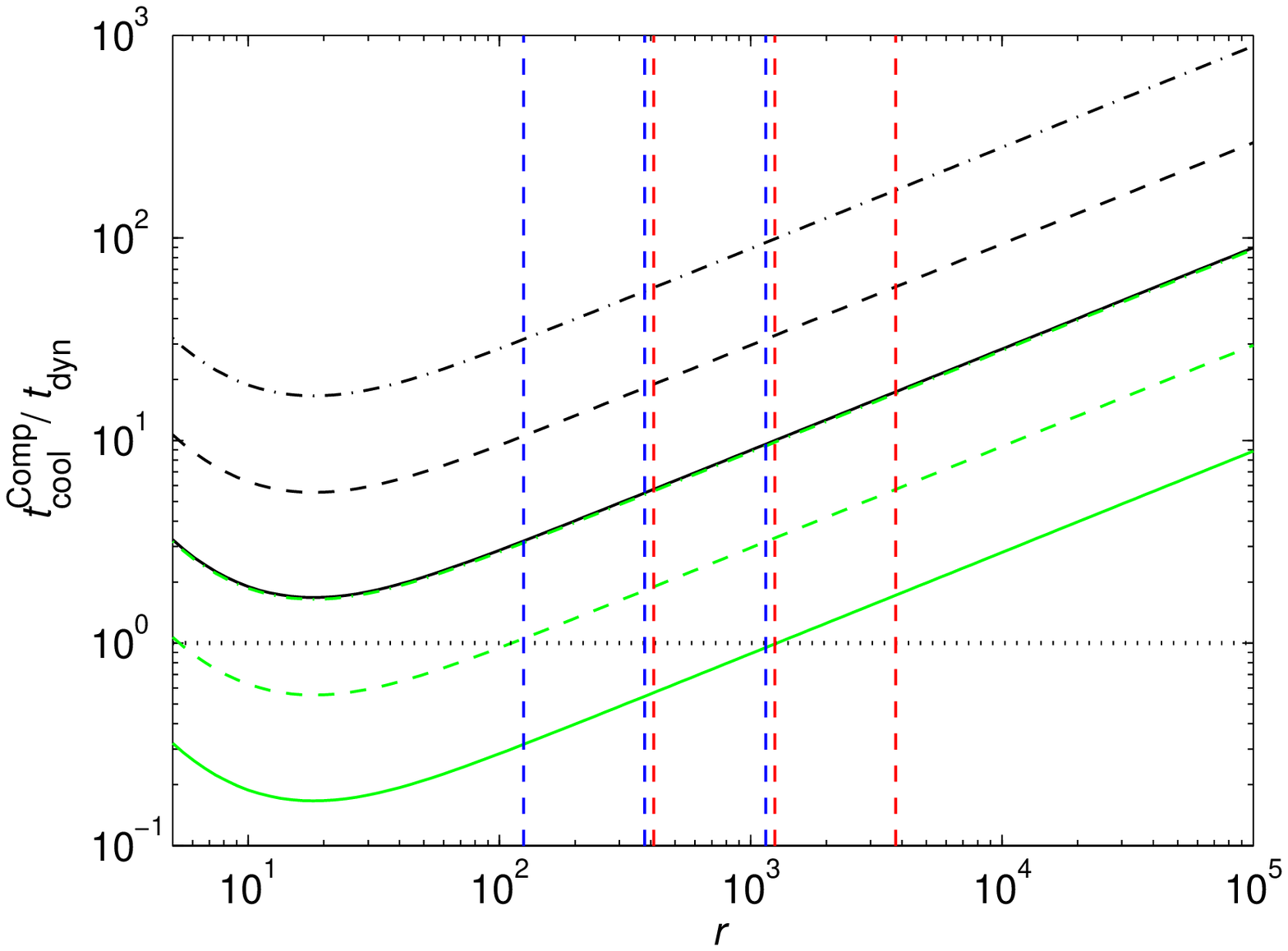}}
\figcaption{The ratios of the Compton cooling timescale to the
dynamical timescale as functions of radius $r$ in the presence of a
small cold accretion disk co-existing with an ADAF in the inner
region.  The black lines represent the ratios calculated with
$\xi_{\rm e}=0.1$ for different Eddington ratios of the small cold
accretion disk: $\lambda_{\rm sd}=0.01$ (solid), $0.003$ (dashed),
and $0.001$ (dotted), respectively. The green lines are the same as
those black lines, but $\xi_{\rm e}=1$ is adopted. All other lines
have the same meanings as those in Fig. \ref{fig_comp}.
\label{fig_comp_smd} }\centerline{}

For the LLAGNs, the radius of the BLR should be lower than that for
broad-line AGNs, if the correlation between $R_{\rm BLR}$ and
$L_{\rm bol}$ (Eq. \ref{r_blr}) still holds for low-luminosity
sources \citep*[but also see][]{2003MNRAS.340..793W}. Our estimate
shows that the radiative cooling of the outflow in the source
accreting at a rate significantly lower than $\dot{m}_{\rm crit}$ is
inefficient, which means that the outflow being expanding
adiabatically is a good approximation. Considering a small volume
$V$ in the outflow with gas temperature $T_{\rm gas}$ and particle
number density $n$, we have
\begin{equation}
dUV={\frac {3}{2}}dp_{\rm gas}V=-p_{\rm gas}dV, \label{energy}
\end{equation}
for an adiabatic expanding outflow, where $p_{\rm gas}=nkT_{\rm
gas}$. The conservation of particles requires
\begin{equation}
{\frac {dV}{V}}=-{\frac {dn}{n}}.\label{conserv}
\end{equation}
Substituting Eq. (\ref{conserv}) into (\ref{energy}), we arrive at
\begin{equation}
d\ln T_{\rm gas}={\frac {2}{3}}d\ln n, \label{tgas_n}
\end{equation}
i.e., $T_{\rm gas}\propto n^{2/3}$. As the number density $n\propto
r^{-3/2}$ in the outflow (see Eq. \ref{dens_w2}), we find that the
gas temperature $T_{\rm gas}\propto r^{-1}$ in an adiabatically
expanding outflow.

\section{Discussion}



The broad-line AGNs are relatively luminous, which contain cold
accretion disks. The accretion flows transit to hot ADAFs when the
sources are accreting at very low rates. Strong outflows may
probably be present in LLAGNs, as the ADAFs have a positive
Bernoulli parameter \citep{1995ApJ...444..231N}. This implies that
the disappearance of BLR in LLAGNs cannot be simply attributed to
the lack of outflows from the accretion disk.

We estimate the cooling of the hot outflows from the ADAF, and find
that the radiative cooling of the outflows is always inefficient
within the radius of the BLR with any values of the parameters
adopted (see Fig. \ref{fig_comp}). The internal energy $U\propto
n_{\rm e}$, and the cooling rate $F^{-}\propto n_{\rm e}^2$, which
indicates that the cooling timescale increases with decreasing
electron number density $n_{\rm e}$. In the estimate of the cooling,
we assume that the radial velocity of the outflow is the same as the
virialized velocity, which is the least velocity that the outflow
can escape to infinity. If the gases in the outflow move at the
speed higher than the virialized velocity, the number density
$n_{\rm e}$ of the electrons decreases with increasing outflow
velocity provided all other parameters are fixed, and therefore the
cooling timescale becomes larger for higher outflow velocity. The
results plotted in Fig. \ref{fig_r_blr} are calculated with
$\eta_{\rm w}=1$, i.e., $\dot{M}_{\rm w}=\dot{M}$, and
$\dot{m}=\dot{m}_{\rm crit}=0.01$, which, of course, leads to an
lower limit on the cooling length scale (see Eqs. \ref{l_cool_min}
and \ref{l_cool_min3}). For most of the LLAGNs, the two parameters,
$\eta_{\rm w}\ll 1$ and $\dot{m}\ll \dot{m}_{\rm crit}$, are
satisfied, which strengthens the conclusion derived in our
estimates.

The detailed physics for the transition of accretion modes is still
unclear. It was suggested that the ADAF co-exists with the standard
disk, i.e., the inner ADAF connects to the outer thin accretion
disk, in some sources accreting at rates slightly lower than the
critical rate $\dot{m}_{\rm crit}$
\citep*[e.g.,][]{1999ApJ...525L..89Q,2003ApJ...599..147C,2009RAA.....9..401X}.
The transition radius increases with decreasing accretion rate
$\dot{m}$, which is expected by the thermal instability or disk
evaporation induced transition scenarios
\citep*[e.g.,][]{1995ApJ...438L..37A,1999ApJ...527L..17L,2000A&A...360.1170R,2002A&A...387..918S}.
In the presence of an outer cold disk, the soft photons from the
cold disk will be Compton upscattered by the hot electrons in the
outflow. {For the ADAF accreting at a rate lower than $\dot{m}_{\rm
crit}$ but with $L_{\rm bol}/L_{\rm Edd}\ga 10^{-3}$, a secondary
inner cold small disk will surround the black hole together with an
ADAF. The mass accretion rate of the small cold disk is regulated by
the condensation process, which is always significantly lower than
the total accretion rate \citep*[see][for the
details]{2007ApJ...671..695L}. Similar to the accretion disk-corona
system, the small cold disk is also irradiated by the ADAF, which
implies that the luminosity of the small disk should be less than a
half of the bolometric luminosity. We adopt $\xi_{\rm e}=0.1$ in our
calculations of the Compton cooling in the outflow near the ADAF,
while $\xi_{\rm e}=1$ is adopted in the calculations for the outflow
far from the ADAF. We find that the Compton cooling of the outflow
near the ADAF is always inefficient due to the soft seed photons
from the outer cold disk (see Fig. \ref{fig_comp}). The situation is
similar for the small inner cold disk, even if the luminosity of the
inner cold disk is as high as $L_{\rm sd}=0.01L_{\rm Edd}$ (see Fig.
\ref{fig_comp_smd}). In the region of the outflow with large
distances from the ADAF, the electrons may have the same temperature
as the ions, i.e., $\xi_{\rm e}=1$. In this case, our results show
that the outflow can be Compton cooled efficiently at large
distances, provided the transition radius of the ADAF to the outer
cold disk is small ($r_{\rm d,tr}\la 20$) or/and the secondary small
cold disk has a luminosity $L_{\rm sd}\ga 0.003L_{\rm Edd}$. We note
that our estimates of the importance of the Compton cooling are
independent of the density of the outflow, i.e., the mass loss rate
in the outflow, which is due to both the Compton cooling rate and
the internal energy of the gas being proportional to the density of
the gas in the outflow. The cold outflows can still be driven from
the outer cold thin disk if the sources are accreting at rates
slightly lower than $\dot{m}_{\rm crit}$, i.e., the transition
radius is not very large. In this case, the outflow from the ADAF
can still be cooled at large distances from the black hole due to
the Compton scattering of the soft seed photons from the outer cold
disk or/and the secondary small inner cold disk. The small inner
cold disk is evaporated completely in the ADAF, which may connect to
the outer thin disk at a very large radius (or the outer cold disk
is suppressed by the ADAF), when $L_{\rm bol}/L_{\rm
Edd}\la10^{-3}$, and therefore the BLR disappears due to the lack of
cold outflow from the disk or the cooling of the outflow from the
ADAF being inefficient. This is consistent with the observations
that almost all ``true" type 2 AGNs have mass accretion rates
$L_{\rm bol}/L_{\rm Edd}\la10^{-3}$
\citep*[e.g.,][]{2003ApJ...589L..13N}.}

For the cases that the radiatively cooling can be neglected, the
temperature of the gas will drop in an adiabatically expanding
outflow. Our estimate shows that the gas temperature $T_{\rm
gas}\propto r^{-1}$ in the outflow. The typical temperature of the
ions in an ADAF near the black hole is $\sim 10^{11-12}$~K
\citep*[e.g.][]{2008NewAR..51..733N}, the gases can be cooled to the
typical temperature of BLRs ($\sim 10^4$~K) only in the outflow with
a distance $>10^{6-7}$ Schwarzschild radii from the black hole. It
corresponds to $\sim 10^{4-5}$ light days for a black hole with
$M=10^{7}M_\odot$, which is obviously beyond the BLR in luminous
AGNs (see Fig. \ref{fig_r_blr}). Therefore, we propose that the
outflows from the ADAFs in LLAGNs are too hot to be cooled to form
clouds in the BLRs, which leads to the disappearance of the BLR in
LLAGNs.

A small fraction of AGNs were found to have emission lines with
double-peaked profiles
\citep*[e.g.,][]{1994ApJS...90....1E,2003AJ....126.1720S}, which
usually have low Eddington ratios \citep*[see][for a review and
references therein, but also see Wu \& Liu 2004; Bian et al.
2007]{2006ASPC..360..217E}. The most favorite model for the
double-peaked emitters suggests that the double-peaked broad
emission lines are emitted from a ring in the accretion disk, which
may also be photo-ionized by the radiation from the inner region
or/and the outflow
\citep*[e.g.,][]{1989ApJ...339..742C,2006ApJ...643..652N,2006ApJ...652..112C}.
The observed broad-line emission may originate from two separated
regions: the clouds in the normal BLRs, or/and the outer ring in the
thin accretion disk. The broad-line emission from the BLR clouds
dominates over that from the outer region of the accretion disk in
normal broad-line AGNs. For the double-peaked emitters accreting at
rates lower than the critical accretion rate $\dot{m}_{\rm crit}$,
the ADAF is present in the inner region and connects to the outer
thin accretion disk. The gases in the outflow from the ADAF are too
hot to be cooled to form the clouds in the BLR when the transition
radius of the ADAF to the outer disk $r_{\rm d,tr}\ga20$ and the
secondary small cold disk is less luminous than $L_{\rm sd}\la
0.003L_{\rm Edd}$, which leads to the disappearance of BLR clouds in
these sources. Thus, the line emission from the outer region of the
accretion disk is not contaminated by the emission from the BLR
clouds, which emerges as double-peaked emission lines. This also
provides a clue to the theoretical models for the accretion mode
transition.

\acknowledgments I thank the referee for the very helpful
comments/suggestions. This work is supported by the NSFC (grants
10773020, 10821302, and 10833002), the National Basic Research
Program of China (grant 2009CB824800), the Science and Technology
Commission of Shanghai Municipality (10XD1405000), the CAS (grant
KJCX2-YW-T03), and the CAS/SAFEA International Partnership Program
for Creative Research Teams.

{}

\end{document}